\documentclass[12pt]{article}

\usepackage{cite}
\usepackage{epsf}
\usepackage{graphicx}
\usepackage{psfrag}
\usepackage[dvips]{epsfig}

\usepackage[english]{babel}

\usepackage{amssymb,indentfirst}
\usepackage{amsmath}

\usepackage{verbatim}

\makeatletter
\renewcommand \thesection {\@arabic\c@section.}
\renewcommand\thesubsection   {\thesection\@arabic\c@subsection.}
\renewcommand\thesubsubsection{\thesubsection\@arabic\c@subsubsection.}
\makeatother

\def\starup#1{\mbox{$\raise1.8ex\hbox{$*$} \kern-.7em#1$}}
\def\krup#1{\mbox{$\raise1.8ex\hbox{$+$} \kern-1.0em#1$}}
\def\linup#1{\mbox{$\raise1.9ex\hbox{---} \kern-1.0em#1$}}

\begin{document}
\title{On mass limit for chiral color symmetry $G'$-boson
\\ from Tevatron data on $t \bar{t}$ production
}

\author{M.V.~Martynov\footnote{E-mail: martmix@mail.ru}, \,
 A.D.~Smirnov\footnote{E-mail: asmirnov@univ.uniyar.ac.ru}\\
{\small Division of Theoretical Physics, Department of Physics,}\\
{\small Yaroslavl State University, Sovietskaya 14,}\\
{\small 150000 Yaroslavl, Russia.}}
\date{}
\maketitle


\begin{abstract}
\noindent
The contributions of $G'$-boson predicted by the chiral color symmetry of quarks
to the cross section $\sigma_{t\bar{t}}$ and to the forward-backward asymmetry
$A_{\rm FB}^{p \bar p}$ of $t\bar{t}$ production at the Tevatron
are calculated with account of the difference of the strengths
of the $\bar q q G$ and $\bar q q G'$ interactions.
The results are analysed in dependence on two free parameters of the model,
the mixing angle $\theta_G$ and $G'$ mass $m_{G'}$.
The $G'$-boson contributions to $\sigma_{t\bar{t}}$ and $A_{\rm FB}^{p \bar p}$ are shown
to be consistent with the Tevatron data on $\sigma_{t\bar{t}}$ and $A_{\rm FB}^{p \bar p}$, 
the allowed region in the $m_{G'} - \theta_G$ plane is discussed 
and around  $m_{G'}=1.2 \, TeV, \; \theta_G=14^\circ $ 
the region of~$1 \sigma$ consistency is found.

\vspace{5mm}


\end{abstract}




The Standard Model (SM) of electroweak and strong interactions based on the gauge symmetry group
\begin{eqnarray}
G_{SM}=SU_c(3) \times SU_L(2) \times U(1)
\label{eq:GSM}
\end{eqnarray}
\noindent
is now the reliable theoretical basis for description of interactions
of guarks and leptons and gauge fields at the energies of order of hundreds GeV.
At the same time the SM leaves some questions within itself open and seems to be
only a first step in our understanding the fundamental interactions at more high energies
and the investigations of the possible extentions of the SM is one of the aims of
the modern elementary particle physics.
The simplest extentions of the SM (such as two Higgs models, models based on supersymmetry,
left-right symmetry, four color quark-lepton symmetry or models implying the four fermion generation, etc.)
predict the new physics effects at one or a few TeV energies and are most interesting now in anticipation
of the new results from the LHC which will allow the investigations of new physics effects
at the TeV energy scale with very large statistics \cite{Butterworth:2007bi}.

One of such simple extentions of the SM can be based on the idea of the originally chiral character
of $SU_c(3)$ color symmstry of quarks. i.e on the gauge group of the chiral color symmetry
\begin{equation}
\label{chiral_group}
G_c=SU_L(3)\!\times \! SU_R(3) \to SU_c(3),
\end{equation}
which is assumed to be valid at high energies and is broken to usual QCD $SU_c(3)$ at low energy scale.
Such chiral color theories~\cite{Pati:1975ze,Hall:1985wz,Frampton:1987ut,Frampton:1987dn}
in addition to the usual massless gluon $G_\mu$ predict in the simplest case of $g_L=g_R$
the existence of a new color-octet gauge boson, the axigluon $G^A_\mu$ with mass $m_{G_A}$.
The axigluon couples to quarks with an axial vector structure with the strong interaction
coupling constant $\alpha_s$ and has a width $\Gamma_{G_A}\approx 0.1 m_{G_A}$~\cite{Bagger:1988}.
Since it is the colored gauge particle with axial vector coupling to quarks, the axigluon should
give rise the increase of the hadronic cross section as well as the appearance 
of a spin-1 resonance in $t\bar{t}$ invariant mass distribution\cite{frederix-2009} and   
of a forward-backward asymmetry of order $\alpha_s^2$ \cite{rodrigo-2008} in the $t\bar{t}$ production.
The lower mass limit for the axigluon from the Tevatron data have been found
in ref.~\cite{rodrigo-2008,antunano-2007}.
The massive color octet with arbitrary vector-- and axial-vector--quark coupling constants
has been considered phenomenologicaly with analysys of mass limits
for this octet in dependence on its coupling constants in ref.~\cite{ferrario-2008}.

The color octet as the gauge boson $G'$ induced by the chiral color symmetry
in general case of~$g_L\neq g_R$ was considered firstly in ref.~\cite{Cuypers:1990hb} and then
with analysys of its phenomenology at the Tevatron and LHC in ref.~\cite{Martynov:2009en}.
The model has two free parameters, $G'$-boson mass $m_{G'}$ and the $G^{L} - G^{R}$ mixing angle $\theta_G$,
$tg\,\theta_G=g_R/g_L$.
Using the CDF data on cross section~\cite{Lister:2008it}
and forward-backward asymmetry~\cite{aaltonen-2008} of the $t\bar{t}$ production at the Tevatron
\begin{eqnarray}
\sigma_{t\bar{t}} & = & 7.0 \pm 0.3 (stat) \pm 0.4 (syst) \pm 0.4 (lumi) pb ,
\label{expcspptt}
\\
    A_{\rm FB}^{p \bar p} & = &
0.17 \pm 0.07~(\rm{stat}) \pm 0.04~(\rm{sys})
\label{AFBpptt}
\end{eqnarray}
the $m_{G'} - \theta_G$ region simultaneusly
compatible with data~\eqref{expcspptt}~and~\eqref{AFBpptt}  within $2 \sigma$ was 
found~\cite{Martynov:2009en}
with asuming the same values of $\alpha_s$ in $\bar q q G$ and $\bar q q G'$ interactions.

The SM predictions for $\sigma_{t\bar{t}}$ and $A_{\rm FB}^{p \bar p}$ have been discussed in 
refs.~\cite{Cacciari:2008zb,Kidonakis:2008mu,Moch:2008ai} 
and~\cite{Kuhn:1998kw,Bowen:2005ap,antunano-2007,Almeida:2008ug} 
respectively and we quote here the next SM predictions for $\sigma_{t\bar{t}}$~\cite{Cacciari:2008zb} 
and $A_{\rm FB}^{p \bar p}$~\cite{antunano-2007}    
\begin{eqnarray}
\sigma_{t\bar{t}}^{SM} & = & 7.35 {~}^{+0.38}_{-0.80}~\mathrm{(scale)}
{~}^{+0.49}_{-0.34} ~\mathrm{(PDFs)} \mathrm{[CTEQ6.5]}  \, \mathrm{pb}  \div  
\label{sppttSM}
\\ 
\notag &\phantom{\div}& 7.93 {~}^{+0.34}_{-0.56}~\mathrm{(scale)}
{~}^{+0.24}_{-0.20} ~\mathrm{(PDFs)} \mathrm{[MRST2006nnlo]} \, \mathrm{pb} ,
\\
  A_{\rm FB}^{SM}(p \bar p \to t \bar t )  &=&  0.051(6). 
\label{AFBppttSM}
\end{eqnarray}
The first and second values in~\eqref{sppttSM} were obtained in  
NLO+NLL approximation with $m_t=171$~GeV and correspond to the different 
choises of the parton distribution functions (CTEQ6.5 and MRST2006nnlo respectively).     
As seen the experimental and theoretical values of $\sigma_{t\bar{t}}$~\eqref{expcspptt},~\eqref{sppttSM}  
are compatible within the experimental and theoretical errors 
whereas the experimental value of $A_{\rm FB}^{p \bar p}$~\eqref{AFBpptt} exceeds 
the corresponding theoretical prediction~\eqref{AFBppttSM}.  
This deviation is not so large with account of the experimental errors nevertheless this circumstance induces 
the discussion of this situation in the literature~\cite{Ferrario:2009bz, Zerwekh:2009vi, 
Frampton:2009rk,Shu:2009xf,Ferrario:2009ee,Cao:2010zb, Dorsner:2009mq,Jung:2009jz,Djouadi:2009nb,
Cheung:2009ch,Arhrib:2009hu,Barger:2010mw,Cao:2009uz}.
       
At the present time the CDF data on cross section~\cite{CDF9913}
and forward-backward asymmetry~\cite{CDFAFB2009} of the $t\bar{t}$ production at the Tevatron are updated as
\begin{eqnarray}
\sigma_{t\bar{t}} & = & 7.5 \pm 0.31 (stat) \pm 0.34 (syst) \pm 0.15 (lumi) pb \, (= 7.5 \pm 0.48 \, pb) ,
\label{expcspptt09}
\\
A_{\rm FB}^{p \bar p} & = & 0.193 \pm 0.065~(\rm{stat}) \pm 0.024~(\rm{sys})\, (= 0.193 \pm 0.069)
\label{AFBpptt09}
\end{eqnarray}
so that the difference between the central value in~\eqref{AFBpptt09} and
the SM prediction~\eqref{AFBppttSM} exceeds now~$2 \sigma$ and the gauge $G'$-boson~\cite{Martynov:2009en}
could look as disfavoured~\cite{Ferrario:2009bz}.
In this situation it is reasonable to analyse the contributions of the gauge $G'$-boson
into cross section and forward-backward asymmetry of the $t\bar{t}$ production at the Tevatron
more carefully, with the more correct account of the strength of the $\bar q q G'$ interaction.

In the present paper we calculate the contributions of the gauge $G'$-boson to the cross section
and to the forward-backward asymmetry of the $Q \bar{Q}$ production in $p \bar{p}$ collisions
with account of the running coupling constant $\alpha_s(M_{chc})$ at the mass scale of
the chiral color symmetry breaking $M_{chc}$.
We compare the results with the Tevatron data~\eqref{expcspptt09},~\eqref{AFBpptt09}
and discuss the corresponding allowed $m_{G'} - \theta_G$ region of the free parameters of the model.

The interaction of the $G'$-boson with quarks can be written as
\begin{equation}
\mathcal{L}_{G'qq}=g_{st}(M_{chc}) \, \bar{q} \gamma^\mu (v + a \gamma_5) G'_\mu q
\end{equation}
where $g_{st}(M_{chc})$ is the strong interaction coupling constant 
at the mass scale $M_{chc}$ of the chiral color symmetry breaking 
and $v$ and $a$ are the phenomenological vector and axial-vector coupling constants.
The gauge symmetry~\eqref{chiral_group}  gives for $v$ and $a$ the expressions
\begin{equation}
v = \frac{c_G^2-s_G^2}{2 s_G c_G} = \cot(2\theta_G), \,\,\,\,
a = \frac{1}{2 s_G c_G} = 1 / \sin(2\theta_G) .
\label{eg:va}
\end{equation}
where
\begin{eqnarray}
s_G =\sin\theta_G = \frac{g_R}{\sqrt{(g_L)^2+(g_R)^2}},  \,\,\,\,
c_G =\cos\theta_G = \frac{g_L}{\sqrt{(g_L)^2+(g_R)^2}},
\label{eq:sGcG}
\end{eqnarray}
$\theta_G$ is $G^{L} - G^{R}$ mixing angle, $g_L$ and $g_R$ are the gauge coupling constants
satisfying the relation
\begin{eqnarray}
\frac{g_L g_R}{\sqrt{(g_L)^2+(g_R)^2}} = g_{st}(M_{chc}). 
\label{eq:gLgRgst}
\end{eqnarray}

The differential partonic cross section of the process $q\bar{q} \rightarrow Q \bar{Q}$
considering the $G'$-boson and gluon contributions within the tree approximation
with account of the difference of the strengths of the $\bar q q G$ and  $\bar q q G'$ interactions
has the form
\begin{eqnarray}
\nonumber
&&\frac{ d\sigma(q\bar{q} \stackrel{\,g,\,G'}{\rightarrow} Q \bar{Q}) }{d\cos \hat{\theta}} = 
 \frac{\pi \beta}{9\hat{s}}
\bigg \lbrace \alpha_s^2(\mu) \, f^{(+)} +
\frac{\alpha_s(\mu) \, \alpha_s(M_{chc}) \, 2 \hat{s} (\hat{s}-m_{G'}^2)}
{(\hat{s}-m_{G'}^2)^2+m_{G'}^2 \Gamma_{G'}^2}
\Big[ \, v^2 f^{(+)} + 2 a^2 \beta c \, \Big] +
\\ \label{diffsect}
&& + \frac{\alpha_s^2(M_{chc}) \, \hat{s}^2} {(\hat{s}-m_{G'}^2)^2+m_{G'}^2 \Gamma_{G'}^2}
\Big[ \left( v^2 + a^2 \right)
\big( v^2 f^{(+)}+   a^2 f^{(-)} \big)
+ 8 a^2v^2 \beta c \, \Big]
\bigg \rbrace,
\end{eqnarray}
where $f^{(\pm)}=(1+\beta^2 c^2\pm 4m_Q^2/\hat{s})$, $c = \cos \hat{\theta}$,
$\hat{\theta}$ is the  scattering angle of $Q$-quark in the parton center of mass frame,
$\hat{s}$ is the invariant mass of $Q \bar{Q}$ system,
$\beta = \sqrt{1-4m_Q^2/\hat{s}}$, $M_{chc}$ is  the mass scale of the chiral color symmetry breaking
and $\mu$ is a typical scale of the process.

The corresponding to~\eqref{diffsect} total cross section takes the form
\begin{eqnarray}
\nonumber
\sigma(q\bar{q} \stackrel{\,g,\,G'}{\rightarrow} Q \bar{Q}) &=& \frac{4\pi \beta}{27\hat{s}}
\bigg \lbrace
\alpha_s^2(\mu) \, (3-\beta^2) +
\frac{2\, \alpha_s(\mu)  \alpha_s(M_{chc})\, v^2   \hat{s}(\hat{s}-m_{G'}^2)(3-\beta^2)}
{(\hat{s}-m_{G'}^2)^2+\Gamma_{G'}^2 m_{G'}^2}+\\
&+&\frac{\alpha_s^2(M_{chc}) \, \hat{s}^2 \big[ \, v^4(3-\beta^2) + v^2 a^2 (3+\beta^2) +
2a^4\beta^2 \, \big]}
{(\hat{s}-m_{G'}^2)^2 + \Gamma_{G'}^2 m_{G'}^2}
\bigg \rbrace.
 \label{sect}
\end{eqnarray}

The enterring into~\eqref{diffsect},~\eqref{sect} hadronic width $\Gamma_{G'}$ of the $G'$-boson
is known~\cite{ferrario-2008,Martynov:2009en}. Its magnitude is proportional to $ \alpha_s = \alpha_s(M_{chc})$
and for example at $M_{chc}=1.2 \, TeV$ we obtain the next estimations for the relative width of $G'$-boson
\begin{equation}
  \Gamma_{G'}/m_{G'}=0.08, \; 0.14, \; 0.33, \; 0.60, \; 1.37
\label{gammaG1}
\end{equation}
for  $ \theta_G=45^\circ, \; 30^\circ, \; 20^\circ, \; 15^\circ, \; 10^\circ $
respectively.

As is known the $G'$-boson in tree approximation does not contribute to the partonic process
$g g  \rightarrow Q \bar{Q}$ of gluon fusion and the differential and total SM cross sections
of this process are well known and defined by $\alpha_s(\mu)$.

The $G'$-boson can generate, at tree-level, a forward-backward asymmetry through the interference
of $q\bar{q} \stackrel{\,G'}{\rightarrow} t\bar{t}$ and $q\bar{q} \stackrel{\,g}{\rightarrow} t\bar{t}$
amplitudes~\cite{antunano-2007,Sehgal:1987wi,choudhury-2007}.
From~\eqref{diffsect} we obtain the forward-backward difference
in the $q\bar{q}\to Q\bar{Q}$ cross section in the form
\begin{eqnarray}
\Delta_{FB}(q\bar{q}\to Q\bar{Q})&=& \sigma(q\bar{q}\rightarrow Q \bar{Q}, \, \cos \theta > 0)-
\sigma(q\bar{q}\rightarrow Q \bar{Q}, \, \cos \theta < 0)=
\nonumber
\\
&=&
\frac{4\pi \beta^2 a^2}{9}
\Bigg( \frac{\alpha_s(\mu) \, \alpha_s(M_{chc}) \, (\hat{s}-m_{G'}^2) + 2 \alpha_s^2(M_{chc}) \, v^2 \hat{s}}
{(\hat{s}-m_{G'}^2)^2+m_{G'}^2 \Gamma_{G'}^2}
\Bigg)
\label{eq:deltaFBqq}
\end{eqnarray}
which can give rise to the corresponding forward-backward asymmetry $A_{\rm FB}^{p \bar p}$
of $t\bar{t}$-pair production in $p \bar p$ collisions at the Tevatron. 
Notice that the first term in~\eqref{diffsect} describing the SM contribution in 
the differential parton cross section in tree approximation does not contribute 
in the forward-backward difference~\eqref{eq:deltaFBqq} and in the forward-backward 
asymmetry $A_{\rm FB}^{p \bar p}$ as well.

We have calculated the cross section $\sigma(p \bar{p} \rightarrow t \bar{t})$
of $t\bar{t}$-pair production in $p\bar{p}$-collisions at the Tevatron energy
using the total parton cross section of quark-antiquark annihilation~\eqref{sect},  
the total SM parton cross section of the gluon fusion $g g  \rightarrow Q \bar{Q}$ 
and the parton densities AL'03~\cite{alekhin} (NLO, fixed-flavor-number,  $Q^2=m_t^2$)
with the appropriate K-factor $K=1.24$~\cite{campbell-2007-70}.
Here and below we beleive $\mu^2=Q^2$, $M_{chc}=m_{G'}$. 

With the same parton densities we have calculated and analysed the forward-backward 
asymmetry $A_{\rm FB}^{p \bar p}$ in the form  
\begin{eqnarray}
    A_{\rm FB}^{p \bar p} & = & A_{\rm FB}^{G'} + A_{\rm FB}^{SM},  
\label{AFBppttcalc}
\end{eqnarray}
where $A_{\rm FB}^{G'}$ is the corresponding $G'$ boson contribution which has been calculated 
using the differential parton cross section~\eqref{diffsect} 
(one can use also the expression~\eqref{eq:deltaFBqq}) and $A_{\rm FB}^{SM}$ is the SM prediction for 
$A_{\rm FB}^{p \bar p}$ for which we have used the value~\eqref{AFBppttSM} of ref.~\cite{antunano-2007}.  

We have anlysed the results of calculations in dependence on $m_{G'}$ and $\theta_G$ in comparision with
the data~\eqref{expcspptt09},~\eqref{AFBpptt09}.
The Fig.1 shows the regions in the $m_{G'} - \theta_G$ plane which are simultaneusly consistent
with the data~\eqref{expcspptt09} and~\eqref{AFBpptt09} within $1 \sigma$ (dark region),
$2 \sigma$ (grey region) and $3 \sigma$ (light-grey region).
As seen from the Fig.1 for
\begin{equation}
m_{G'}>1.02 \, TeV
\label{eq:mG1limcsFAB}
\end{equation}
in the $m_{G'} - \theta_G$ plane there is the region which is
consistent with the data~\eqref{expcspptt09},~\eqref{AFBpptt09} within~$1 \sigma$.
For example, for the masses
%
\begin{equation}
a) \; m_{G'} = 1.02 \, TeV, \; \; b) \; m_{G'} = 1.2 \, TeV, \; \; c) \; m_{G'} = 1.4 \, TeV
\label{eq:mG1values}
\end{equation}
with the appropriate values of $\theta_G$
($\theta_G=19^\circ, \, \theta_G=14^\circ, \, \theta_G=11^\circ$ respectively, 
these points are marked in Fig.1 by crosses)
we obtain for $\sigma_{t\bar{t}}$, $A_{\rm FB}^{p \bar p}$ the values
\begin{eqnarray}
&& a)  \; \sigma_{t\bar{t}} = 7.98 \, pb, \; \; A_{\rm FB}^{p \bar p} = 0.158 \, (0.107),
\label{pointa}
\\
&& b)  \; \sigma_{t\bar{t}} = 7.61 \, pb, \; \; A_{\rm FB}^{p \bar p} = 0.154 \, (0.103),
\label{pointb}
\\
&& c)  \; \sigma_{t\bar{t}} = 7.57 \, pb, \; \; A_{\rm FB}^{p \bar p} = 0.141 \, (0.090),
\label{pointc}
\end{eqnarray}
which are consistent with the data~\eqref{expcspptt09},~\eqref{AFBpptt09} within~$1 \sigma$ 
(in parentheses we show for comparision the $G'$-boson contributions in $A_{\rm FB}^{p \bar p}$ 
defined by~\eqref{eq:deltaFBqq}, without the SM contribution~\eqref{AFBppttSM}).
So, the $G'$-boson induced by the chiral color symmetry~\eqref{chiral_group}  
in general case of~$g_L\neq g_R$ is consistent with the data~\eqref{expcspptt09},~\eqref{AFBpptt09} 
and can reduce the difference between the experimental and SM values~\eqref{AFBpptt09}, ~\eqref{AFBppttSM} 
of the forward-backward asymmetry $A_{\rm FB}^{p \bar p}$ in the $t\bar{t}$ production at the Tevatron.

In conclusion, we summarize the results found in this work.

The contributions of $G'$-boson predicted by the chiral color symmetry of quarks
to the cross section $\sigma_{t\bar{t}}$ and to the forward-backward asymmetry
$A_{\rm FB}^{p \bar p}$ of $t\bar{t}$ production at the Tevatron
are calculated with account of the difference of the strengths
of the $\bar q q G$ and $\bar q q G'$ interactions by using
the running coupling constants $\alpha_s(\mu)$ and $\alpha_s(M_{chc})$
at the typical mass scale of the process $\mu$
and at the mass scale of the chiral color symmetry breaking~$M_{chc}$ respectively.
The results are analysed in dependence on two free parameters of the model,
the mixing angle $\theta_G$ and $G'$ mass $m_{G'}$, in comparision with the Tevatron data
on $\sigma_{t\bar{t}}$ and $A_{\rm FB}^{p \bar p}$.
The $G'$-boson contributions to $\sigma_{t\bar{t}}$ and $A_{\rm FB}^{p \bar p}$ are shown
to be consistent with these data, the allowed region
in the $m_{G'} - \theta_G$ plane is discussed and 
it is shown that for $m_{G'}>1.02 \, TeV, \; \theta_G<19^\circ $ 
there is the region in the $m_{G'} - \theta_G$ plane 
with~$1 \sigma$ consistency.

The work is supported by the Ministry of Education and Science of Russia  
under state contract No.NK-533P(7) 
of the Federal Programme "Scientific and Pedagogical Personnel of Innovation Russia"  
for 2009-2013 years.


\newpage

%

\newpage

{\Large\bf Figure captions}

\bigskip

\begin{quotation}
\noindent
Fig. 1. The $m_{G'} - \theta_G$ regions consistent with CDF data
on cross section~$\sigma_{t\bar{t}}$  and forward-backward asymmetry~$A_{\rm FB}^{p \bar p}$
in $t \bar{t}$ production within $1 \sigma$ (dark region), $2 \sigma$ (grey region)
and $3 \sigma$ (light-grey region).
\end{quotation}


\newpage
\begin{figure}[htb]
\vspace*{0.5cm}
 \centerline{
\epsfxsize=0.7\textwidth
\epsffile{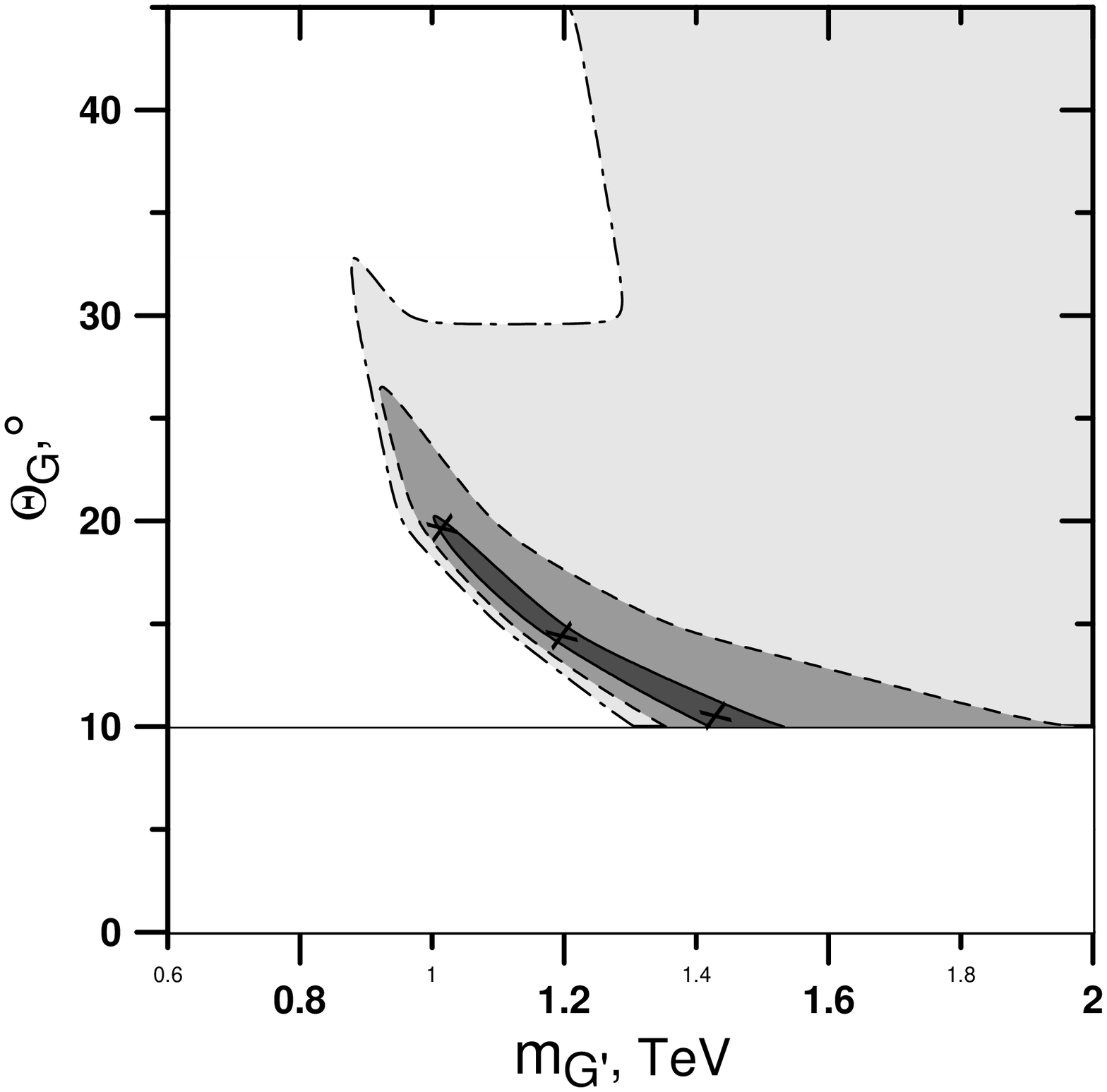}}
\vspace*{1mm}
\caption{}
\label{constrTevatron}
\end{figure}

\vspace*{5cm}
\vfill \centerline{M.V. Martynov, A.D.~Smirnov, Modern Physics Letters A}
\centerline{Fig. 1}


\end{document}